\documentclass[aip,twocolumn,showpacs,floatfix,preprintnumbers,amsmath, amsthm, amssymb]{revtex4}
\usepackage[latin9]{inputenc}
\usepackage{graphicx}
\usepackage{hyperref}
\usepackage{sidecap}
\begin{document}

\title{Electronic, structural, and elastic properties of metal nitrides XN (X = Sc, Y): A first principle study}

\author{Chinedu E. Ekuma$^1$}
\author{Diola Bagayoko$^2$}
\author{Mark Jarrell$^1$}
\author{Juana Moreno$^1$}
\affiliation{$^1$Department of Physics and Astronomy \&
Center for Computation and Technology,
Louisiana State University,
Baton Rouge, LA 70803, USA\\ 
$^2$Department of Physics, Southern University and A $\&$ M 
College, Baton Rouge, LA 70813, USA}

\date{\today}

\begin{abstract}
We utilized a simple, robust, first principle method, based on basis set optimization with the BZW-EF method, to study the electronic 
and related properties of transition metal mono-nitrides: ScN and YN. We solved the KS system of equations self-consistently 
within the linear combination of atomic 
orbitals (LCAO) formalism. It is shown that the band gap and low energy 
conduction bands, as well as elastic and structural properties, can be calculated with a 
reasonable accuracy when the LCAO formalism is used to obtain an optimal basis. Our calculated, indirect electronic 
band gap (E$^\mathrm{\Gamma-X}_g$) is 0.79 (LDA) and 0.88 eV (GGA) 
for ScN. In the case of YN, we predict an indirect band gap (E$^\mathrm{\Gamma-X}_g$) of 
1.09 (LDA) and 1.15 eV (GGA). We also calculated the equilibrium lattice constants, the bulk 
moduli (B$_{o}$), effective masses, and elastic constants for both systems. Our calculated values are 
in excellent agreement with experimental ones where the latter are available.
\end{abstract}


\maketitle

\section{Introduction}
\label{sec:intro}
There has been great interest in transition-metal nitrides in the past several 
decades \cite{Van1991,Yim1974,PhysRevB.65.045204,Brithen2004}. This interest is fueled 
by their many potential, technological applications, including high hardness, 
\cite{Sundgreen1986} high temperature stability \cite{Edgar1972},  
mechanical strength \cite{gall:6034, Gall1_1998,PhysRevB.65.235307}, magnetic \cite{Houari2008},  
and electronic properties that vary from semiconducting to metallic 
phases \cite{Varek1999,Nesl2000,Pandit2011,Constantin2004,Constantin2005,Ranjan2003,
Ranjan2005,Stampfl2001,Lu2007,Cherchab2008,Zerroung2009,Gregoire2009}.  
These unique properties make XN useful refractory materials and 
hard coatings for cutting tools \cite{Varek1999, Nesl2000, Pandit2011} . 

The aforementioned properties of these materials motivated a variety of 
experimental \cite{Edgar1972,gall:6034,Gall1_1998,Varek1999,Nesl2000,Constantin2004,Constantin2005,
Gregoire2009,Chen2007,Wyckoff1963,Moustakas1996,Harbeke1972,Brithen2000} 
as well as theoretical studies \cite{PhysRevB.65.235307,PhysRevB.65.045204,PhysRevB.65.235307,
Takeuchi2003,Houari2008,Pandit2011,Ranjan2003,Ranjan2005,
Stampfl2001,Lu2007,Cherchab2008,saha:073720,Zerroung2009,Maachou2007,
PhysRevB.65.161204,Lambrecht2000,Bouhemadou2007,Duman2006,Amrania2007} 
 with the latter utilizing computational techniques of varying sophistication, ranging from the tight binding 
and the empirical pseudopotential methods to ab-initio density functional theory (DFT).

Even with the vast number of experimental and computational studies already available for XN, a satisfactory description of their 
electronic, transport (effective mass), elastic, and structural properties is still an area of active research. 
Experimentally, ScN is known to be a semiconductor with a band gap in the 
range of 0.90 $\pm$ 0.1 eV to 1.32 $\pm$ 0.3 eV \cite{Brithen2004,gall:6034}, while most DFT calculations utilizing 
local density approximation (LDA) and generalized gradient approximation (GGA) 
potentials found it to be a metal 
\cite{PhysRevB.65.045204,Maachou2007,Qteish2006,Stampfl2001,Monnier1985,Travaglini1986,Tie2007}. 
Recent Green's function quasiparticle \cite{Lambrecht2000,PhysRev.139.A796}, 
exact exchange \cite{Qteish2006,PhysRevB.59.10031} 
and screened exchange \cite{Stampfl2001,PhysRevB.59.7486} calculations have reported 
that ScN is a semiconductor with an indirect gap (E$^\mathrm{\Gamma-X}_g$) in the range 0.54 -- 1.70 eV. For YN, 
few theoretical calculations utilizing various forms of DFT potentials have reported that it is semi-metallic or 
semiconductor with an indirect band gap (E$^\mathrm{\Gamma-X}_g$) of 0.2 -- 0.3 eV \cite{saha:073720,Mancera2003,Soto2008}, 
0.54 eV \cite{Tie2007}, and 0.80 eV \cite{Zerroug2009}.  
We are not aware of any experiments reporting the indirect band gap of YN.

Theoretical computations have had difficulties in predicting 
the correct band gap energy and other related electronic properties of XN from first principle. 
Indeed, the ``band gap problem'' is decades old. 
Several approaches to solve it have been proposed with significant 
successes. Density functional theory plus additional Couloumb interactions (DFT+U) formalism 
\cite{PhysRevB.44.943, PhysRevB.43.7570, Anisimov1997, Madsen2005} has had good 
successes in obtaining correct energy bands and gaps of materials, but can only be applied to 
correlated and localized electrons, e.g., 3$d$ or 4$f$ in transition and rare-earth oxides. 
Hybrid functionals \cite{heyd:219906,heyd:1187,heyd:8207} 
have also been used in attempts to 
improve on the calculated energy bands and band gaps of materials. This approach involves 
a range separation of the exchange energy into some fraction of 
nonlocal Hartree-Fock exchange potential and a fraction of local spin density approximation (LSDA) or 
generalized gradient approximation (GGA) exchange potential. We should note that this range separation is not 
universal. There is always a range separation parameter $\omega$ which varies between 0 and $\infty$.  
While it is reasonably clear that there exists a value
of $\omega$ that gives the correct gap for a given system, 
this $\omega$ is not universal as its value is always adjusted from one system to another \cite{PhysRevB.78.121201,Henderson2011}.   
For example, in HSE06 \cite{heyd:219906,krukau:224106},  
$\omega$ $=$ 0.11$a_{0}^{-1}$ ($a_{0}$ 
is the Bohr radius) and in Perdew-Burke-Ernzerhof (PBEh) global hybrid \cite{janesko:084111}, it is 25 $\%$ 
short-range exact exchange and 75 $\%$ short-range 
PBE exchange. Even though the HSE functional, in most cases, accurately 
reproduces the optical gap in semiconductors, it severely underestimates 
the gap in insulators \cite{Henderson2011,PhysRevB.76.195440} and the band width
in metallic systems is generally too large \cite{Henderson2011,PhysRevB.76.195440,paier:154709,PhysRevB.83.195134},  
The \citet{PhysRevB.47.13164} (EV) GGA and the \citet{PhysRevLett.102.226401} modified 
Becke-Johnson (TB-mBJ) potentials have also provided some improvements to the calculated band gap of materials. For TB-mBJ, while the 
band gaps are considerably improved, the effective masses are severely underestimated. \cite{PhysRevB.83.195134} 
In the case of the EV-GGA potential, the equilibrium lattice constants are far too large as compared to experiment and, as such, 
leads to an unsatisfactory total energy \cite{PhysRevB.50.7279,PhysRevB.81.195217}. 

The theoretical underestimations of band gaps and other energy eigenvalues have been ascribed to the inadequacies of 
density functional potentials for the description of 
ground state electronic properties of XN \cite{saha:073720}.  
Also, other methods \cite{PhysRevB.75.235102,PhysRevB.74.035101} that 
entirely go beyond the density functional theory (DFT) 
do not obtain the correct band gap values of most semiconductors 
without adjustment or fitting parameters \cite{saha:073720}.  
This unsatisfactory situation is a key motivation for our work. In light of the promising technological 
properties of these materials, parameter-free computations could aid significantly in the design and fabrication 
of devices. To this end, we have also investigated the elastic properties of 
ScN and YN within our parameter-free method. In YN, we predict several of these properties for which there are 
yet no experimental data.

In this paper, we utilized a simple and robust approach based on basis set optimization. 
The rest of this paper is organized as follows. After this introduction in 
Section ~\ref{sec:intro}, the computational method and details used in our work are given in 
Section~\ref{sec:formalism}. Section~\ref{sec:results} shows our computed results. 
Calculated electronic structures are given in Subsection~\ref{subsec:electronic}. 
The results for the chemical bonding and structural properties are presented 
and discussed in Subsections ~\ref{subsec:dos} and ~\ref{subsec:structural}. We will conclude with a 
summary in Section ~\ref{sec:summary}.  

\section{Method and Computational Details}
\label{sec:formalism}
\subsection{Method}
We utilized the electronic structure package developed at Ames Laboratory of the 
U.S. Department of Energy (DOE), Ames, Iowa \cite{Harmon1982}.  
For the LDA computations, we used the Ceperley and Alder DFT 
exchange-correlation contribution \cite{Ceperley1980} as parameterized 
by Vosko-Wilk-Nusair \cite{Vosko1980}. We refer to it as the CA-VWN potential. 
The GGA calculations were carried out using the 
Ceperley and Alder DFT exchange correlation contribution \cite{Ceperley1980} as 
parameterized by Perdew and Wang \cite{Perdew1992}. We refer to it as the 
CA-PW potential.

We employed the linear combination of atomic orbitals (LCAO) approach where an unknown wave function, for the solid state, is 
expressed as a linear combination of atomic orbitals of the individual atoms in the system. The radial parts of these 
orbitals are generally exponential or Gaussian functions resulting from self consistent calculations of energy levels of 
the atomic or ionic species that are present in the solid under study.  We use Gaussian functions and refer to that 
rendition of LCAO as the linear combination of Gaussian orbitals (LCGO).

In addition to the use of DFT potentials and of the LCAO formalism, 
our computational approach rests on the Bagayoko, Zhao, and Williams 
\cite{Bagayoko_1998,Zhao1999} method as recently enhanced by the work of Ekuma 
\cite{Ekuma2010,Ekuma2011,Ekumab2011,chen:2414} and Franklin 
\cite{Franklin2012} (BZW-EF). The method heeds the statement by Kohn and 
Sham \cite{Kohn1965} that the equations defining LDA have to be solved self-consistently. 
Upon the selection of an exchange correlation potential, these equations reduce 
to (1) the one giving the ground state charge density in terms 
of the wave functions of the occupied states and (2) 
the Kohn-Sham equation. We describe below the aforementioned method. 

In the BZW-EF method, we begin our 
calculations with the minimum basis set, 
the one that is just large enough to account for all the electrons in the system under study 
(in this case XN). Our self consistent 
calculation with this basis set is followed by another where the basis set is augmented with one additional orbital 
from one of the atoms or ions in the system. The comparison of the occupied energies from Calculations I and II 
generally shows that they are different, with those from Calculation II generally  
lower than their corresponding ones from I. A third, self consistent calculation is performed with a basis set that 
includes that for Calculation II plus another orbital from one of the atoms in the system. This process of augmenting 
the basis set and of carrying out self consistent calculations continues until a calculation, say N, is found to have 
the \textit{same occupied energies} as calculation (N + 1) that follows it, within computational uncertainty of about 50 meV. 
This convergence of the occupied energies identifies the basis set of Calculation N as the optimal one. The optimal 
basis set is thus the smallest one with which the occupied energies verifiably reach their respective minima \cite{Ekuma2011,ibid2011}. 
The resulting eigenvalues and the corresponding eigenfunctions provide the physics description of the system.

The BZW-EF method has been shown to lead to accurate ground state properties of 
many semiconductors: c-InN \cite{Bagayoko2004}, w-InN \cite{Bagayoko2005}, w-CdS 
\cite{Ekuma2011}, c-CdS \cite{ibid2011}, rutile-TiO$_{2}$ \cite{Ekumab2011}, SrTiO$_\mathrm{3}$ \cite{ekuma:012189}, 
AlAs \cite{Jin2006}, GaN, Si, C, RuO$_{2}$ \cite{Zhao1999}, BaTiO$_\mathrm{3}$ \cite{Bagayoko_1998},
and carbon nanotubes \cite{Zhao2004}.  

In calculations with basis set larger than the optimal one, the ground state charge density does not change, nor do 
the Hamiltonian and the eigenvalues of the occupied states.  Consequently, these calculations do not lower any 
occupied energies (as compared to the results obtained with the optimal basis set), even though they generally 
lead to some lower, unoccupied energies \cite{Ekuma2011,Ekuma2010} by virtue of the Rayleigh 
theorem \cite{Courant1965,Gross1988}. 
The Rayleigh theorem can be stated as follows: if 
one solves an eigenvalue equation with two basis sets I and II, with set II larger than I and where I 
is entirely contained in II, then the eigenvalues obtained with set II are lower than or equal to their corresponding 
ones obtained with basis set I (i.e., E$_i^{N + 1}$ $\leq$ E$_i^{N}$, $\forall$ i $\leqslant$N). This 
theorem explains the reasons that some unoccupied energies are lowered when the 
Kohn-Sham equation is solved with basis sets larger than the optimal one. Such a 
lowering of unoccupied energies with 
basis sets larger than the optimal one is fundamentally different from the one that occurs before the occupied 
energies reach their minima. The latter lowering is ascribed to physical interactions 
given that both the charge density and the Hamiltonian change, from one calculation to the next, before one reaches 
the optimal basis set, while the former one is unphysical. 

Important points about this method include the fact that it provides the 
needed variational freedom for the restructuring of the electronic 
cloud in the material under study, as opposed to the case in isolated 
atoms or ions. Indeed, the methodical increase of the basis set 
remedies possible radial, angular symmetry, and size deficiencies 
by the time the optimal basis set is reached. By changing the basis set, 
i.e., the external input to the charge density equation, the method 
solves self-consistently both this equation and the Kohn Sham equation 
in a way that is quintessentially different from input changes 
limited to new expansion coefficients in the iterative 
process \cite{Bagayoko_2008}. The latter input changes cannot 
remedy any significant deficiency of the basis set, i.e., 
in terms of radial orbital, angular symmetry, or the 
total number of functions (the dimension of the Hamiltonian). 
 In light of the Rayleigh 
theorem, a deficiency could be 
over-completeness as well, in single 
trial basis set calculations. 
\textit{The critical importance of the BZW-EF method resides in part in the 
fact that, by its very variational derivation, 
the validity and the related physical content of 
the eigenvalues of the Kohn Sham equation directly 
hinge on the final, self-consistent charge density 
being as close as possible to that of the real 
material under study.}

The thesis work of Williams \cite{Zhao1999} 
showed that the method may not be needed for 
the description of materials established, 
experimentally, to be metals. At least one 
band crosses the Fermi level in a metal and 
the lowest-laying conduction bands reach 
their minima at the same time as the valence 
bands do. Consequently, the basis set 
and variational effect 
\cite{Bagayoko_1998,Zhao1999} that explains 
the lowering of unoccupied energies, while 
the Hamiltonian does not change, does not 
affect these lowest-laying conduction bands 
in metals, even though it lowers some other 
conduction bands. This fact partly explains 
the early success of DFT for the 
description of metals as opposed to semiconductors and insulators. 
In the initial BZW method, we increased the basis set by 
adding orbitals in the order of increasing energies in 
the atomic or ionic species. Recent works by 
Ekuma \cite{Ekuma2010,Ekuma2011,Ekumab2011,chen:2414} and Franklin 
\cite{Franklin2012} led us to the realization that, for the valence states, 
polarization (i.e., $p$, $d$, and $f$ orbitals) have primacy over 
spherical symmetry ($s$ orbital) in diatomic molecules and 
more complex systems with three or more atomic or ionic sites. 
 Hence, while still useful,  adding orbitals in the order of 
increasing energies in atomic or ionic species can be relaxed.

\subsection{Computational Details}
In the ground state, both ScN and YN have the rock-salt crystal structure. We utilized room temperature experimental lattice 
constant of 4.501 \AA{} \cite{Lengauer1988,gall:6034,ICSD2011,gall:5524} and 4.837 \AA{} \cite{ICSD2011} for ScN and YN, respectively. 
Each Sc (or Y) atom is surrounded by six N atoms, thereby providing the
octahedral environment at the Sc (or Y) site, which leads to the
splitting of the degenerate $d$ orbital into t$_\mathrm{2g}$ and e$_\mathrm{g}$ states.
Every Sc (or Y) atom has 12 nearest neighbors Sc (or Y) atoms and 6 next nearest neighbors
Sc (or Y) atoms. Preliminary calculations indicated that scandium and 
yttrium are closer to Sc$^\mathrm{3+}$and Y$^\mathrm{3+}$ than to the neutral Sc and Y, respectively. 
Similarly, nitrogen is N$^\mathrm{3-}$ as 
opposed to the neutral N. Therefore, we first carried out self consistent 
calculations of the electronic properties of X$^\mathrm{3+}$  (X = Sc, Y) and N$^\mathrm{3-}$. 
Atomic orbitals utilized in these calculations, 
for the valence states, are given between parentheses: Sc$^\mathrm{3+}$ (3$s$3$p$3$d$4$s$4$p$4$d$) and 
N$^\mathrm{3-}$ (2$s$2$p$3$s$3$p$) for ScN and Y$^\mathrm{3+}$ (4$s$4$p$4$d$5$s$5$p$) and N$^\mathrm{3-}$ (2$s$2$p$3$s$3$p$) 
for YN. 
Other atomic states with higher binding energies were treated as 
deep core states. In the optimal basis set for the valence states of ScN, (4$p$4$d$) and (3$p$) are 
unoccupied for Sc$^\mathrm{3+}$ and N$^\mathrm{3-}$, 
respectively. For YN, the unoccupied orbitals in the optimal basis set for the valence states are 
($5p$) and (3$p$) for Y$^\mathrm{3+}$ and N$^\mathrm{3-}$, respectively. 
Nevertheless, these orbitals are included in 
the self-consistent LCAO calculations to allow for a reorganization of electronic cloud in the solid environment, 
including polarization.

In the self-consistency calculation, 
both the potential and charge density are also expanded in terms of even tempered Gaussian orbitals 
(Sc: 15, 15, and 13; Y: 15, 15, and 13; N: 17, 17, and 15 for $s$, $p$, and $d$ orbitals, respectively). 
The exponents, $\alpha$, of the Gaussian basis sets range from 0.19 to $10^{5}$. 
The charge fitting error using the Gaussian functions in the atomic calculation was about $10^{-5}$. Since the deep core 
states are fully occupied and are inactive chemically in the materials, the charge densities of the deep core states 
are kept the same as in the free atoms. However, the core states of low binding energy are still allowed to fully relax, 
along with the valence states, in the self consistent calculations. The computational 
error for the valence charge was 1.2 $\times$ $10^{-5}$ and 5.9 $\times$ $10^{-6}$ per valence electron for ScN and YN, respectively. 
The self consistent potential converged to a difference of $10^{-5}$ after about 60 iterations.

The Brillouin zone (BZ) integration for the charge density in the self consistent iterations was based on 28 special 
k points in the irreducible BZ (IBZ). The energy eigenvalues and eigenfunctions were then obtained at 161 special k points in 
the IBZ for the band structure. A total of 152 weighted k points, chosen along the high symmetry lines in the 
IBZ of ScN and YN, respectively, were used to solve for the energy eigenvalues from which the electron densities 
of states (DOS) were calculated using the linear, analytical tetrahedron method \cite{Lehmann1972}. The partial density of states (pDOS) 
and the effective charge of each atom were calculated using the Mulliken charge analysis procedure \cite{Mulliken1955}.
 We also calculated 
the equilibrium lattice constant (a$_{o}$), the bulk modulus (B$_{o}$) and the electron effective masses 
for different directions with respect to the $\Gamma$ point. For the calculation 
of the equilibrium lattice constant, we utilized 
the Murnaghan equation of state \cite{Murnaghan1968,Murnaghan1995}. By applying an appropriate set of strains to the
undeformed unit cell lattice, we calculated the elastic constants from the resulting change
in total energy on the deformation using the strain-energy method. 
For a typical cubic crystal, three independent elastic moduli 
are of importance; they are usually denoted as C$_\mathrm{11}$, C$_\mathrm{12}$, and C$_\mathrm{44}$.

\section{Results and Discussion}
\label{sec:results}
The results of the electronic structure computations are given 
in Figs.~\ref{fig:scn_elec} to ~\ref{fig:yn_elec}. Figure~\ref{fig:xn_bond} shows the contour plot of the 
distribution of the electron charge density while Figure ~\ref{fig:xn_eos}
depicts the total energy curves of XN. We discuss the electronic structure 
and the effective masses in ~\ref{subsec:electronic}. Densities of States, chemical bonding, 
and electron distribution are described in Subsection~\ref{subsec:dos}. 
The structural and elastic properties are presented in Subsections ~\ref{subsec:structural}. 
\subsection{The Electronic Structure, Band gap and Effective mass}
\label{subsec:electronic}
\begin{figure*}[htb]
  \begin{center}
    \begin{minipage}[t]{0.30\linewidth}
      \raisebox{0cm}{{\includegraphics[trim = 0mm 0mm 5mm 7mm,clip,angle=0,totalheight=0.24\textheight,width=2.3in]{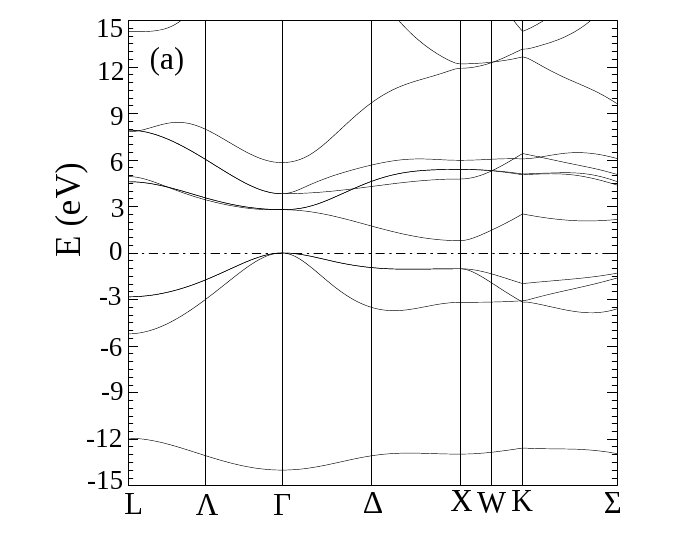}}}
    \end{minipage}\hspace{-0.3cm} 
\hfil
    \begin{minipage}[t]{0.30\linewidth}
      \raisebox{0cm}{{\includegraphics[trim = 0mm 10mm 5mm 8mm,clip,angle=0,totalheight=0.28\textheight,width=2.3in]{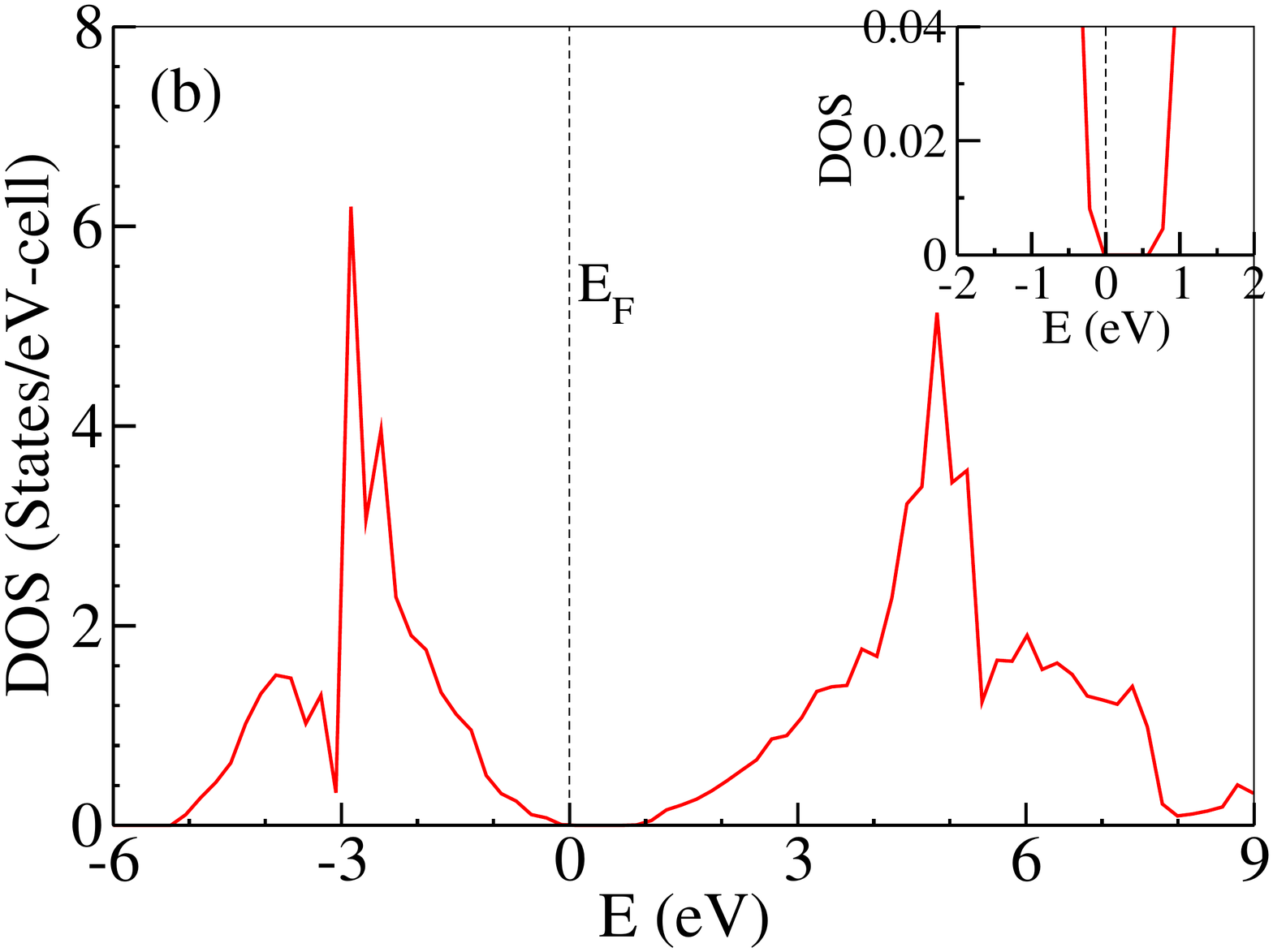}}}
    \end{minipage}
\hfil
    \begin{minipage}[t]{0.30\linewidth}
      \raisebox{0cm}{{\includegraphics[trim = 0mm 10mm 5mm 8mm,clip,angle=0,totalheight=0.28\textheight,width=2.3in]{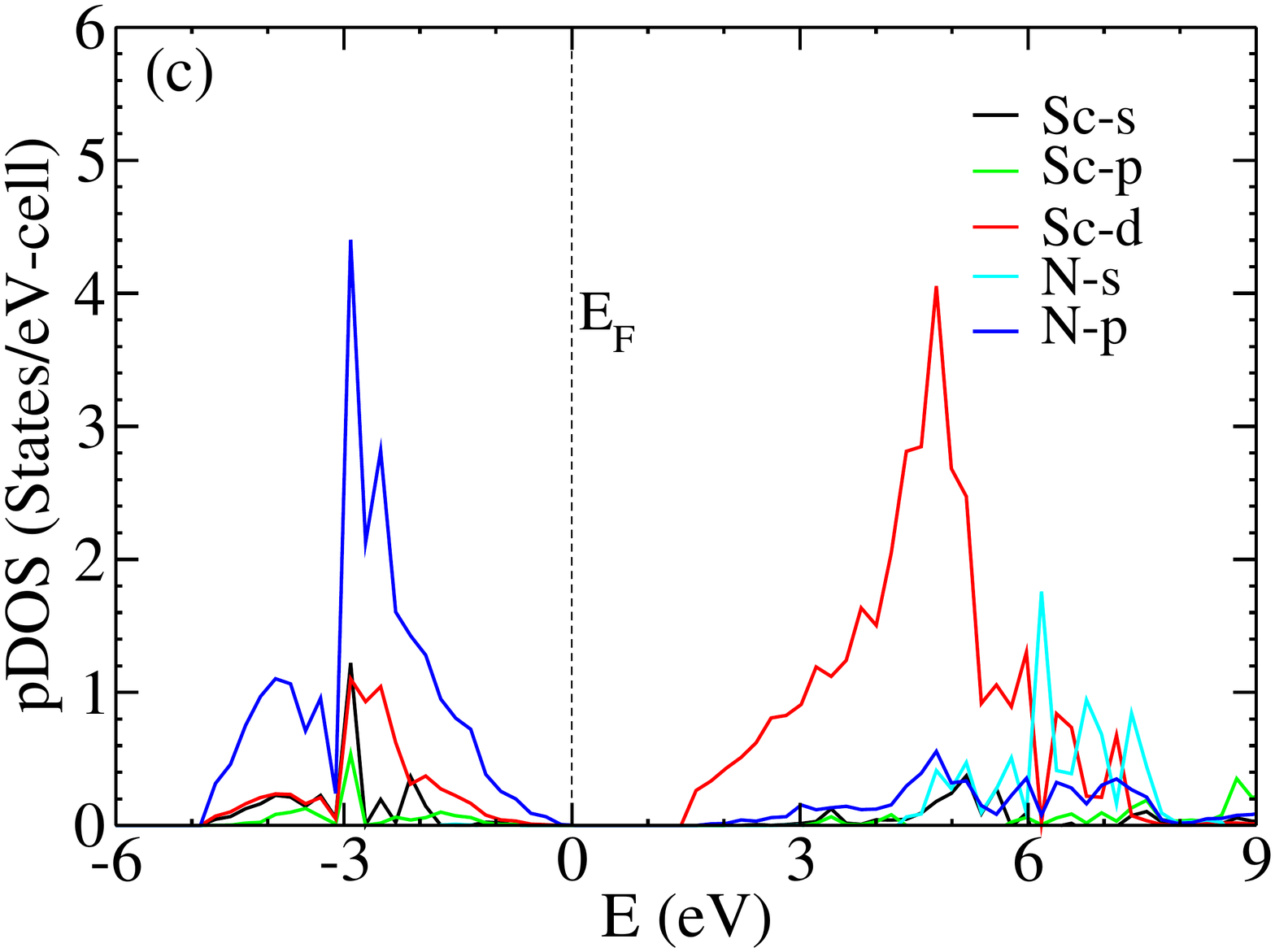}}}
    \end{minipage}
\end{center}
\caption{(Color online). (a) The calculated band structure of c-ScN as obtained with the 
  optimal basis set (for the equilibrium lattice constant of 4.501 \AA{}). 
    The Fermi energy ($\mathrm{E_F}$) has been set equal to zero. 
(b) The calculated density of states (DOS) of c-ScN, 
as obtained from the bands shown in Fig.~\ref{fig:scn_elec}(a). The Fermi energy ($\mathrm{E_F}$) has been set equal to zero. 
(c)  The calculated partial density of states (pDOS) of c-ScN, as obtained 
from the bands shown in Fig.~\ref{fig:scn_elec}(a). The Fermi energy ($\mathrm{E_F}$) has been set equal to zero.}
\label{fig:scn_elec}
\end{figure*}

\begin{figure*}[ht]
  \begin{center}
    \begin{minipage}[t]{0.30\linewidth}
      \raisebox{-5cm}{{\includegraphics[trim = 0mm 0mm 5mm 8mm,clip,angle=0,totalheight=0.24\textheight,width=2.3in]{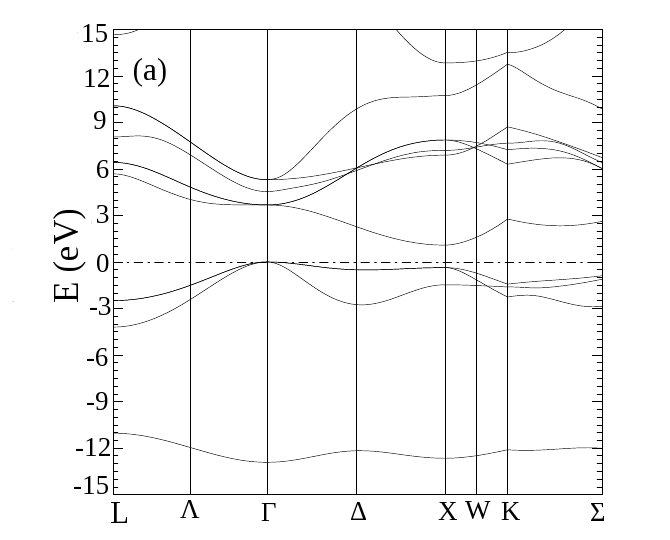}}}
    \end{minipage}\hspace{-0.3cm} 
\hfil
    \begin{minipage}[t]{0.30\linewidth}
      \raisebox{-5cm}{{\includegraphics[trim = 0mm 10mm 5mm 8mm,clip,angle=0,totalheight=0.28\textheight,width=2.3in]{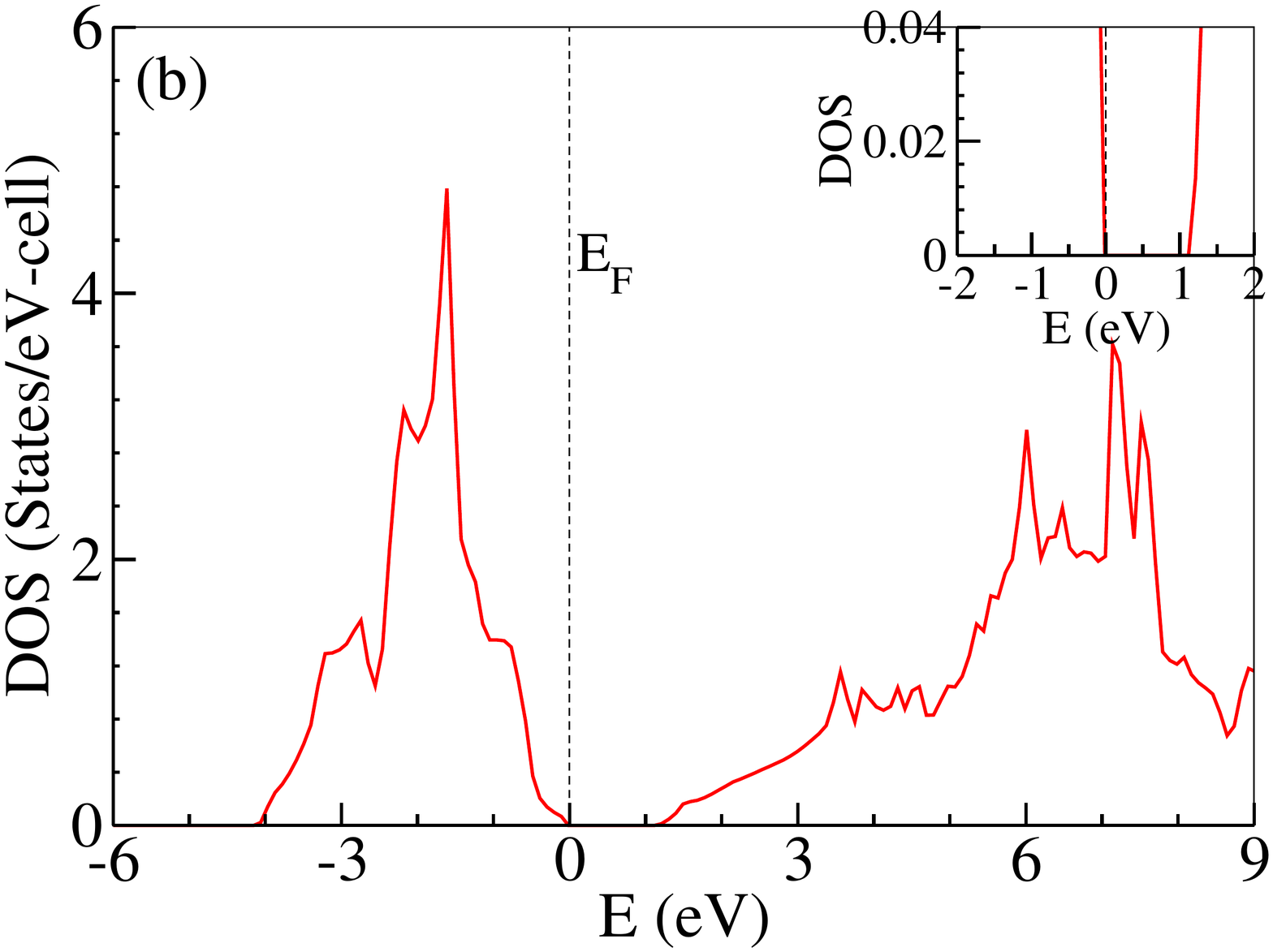}}}
    \end{minipage}
\hfil
    \begin{minipage}[t]{0.30\linewidth}
      \raisebox{-5cm}{{\includegraphics[trim = 0mm 10mm 5mm 8mm,clip,angle=0,totalheight=0.28\textheight,width=2.3in]{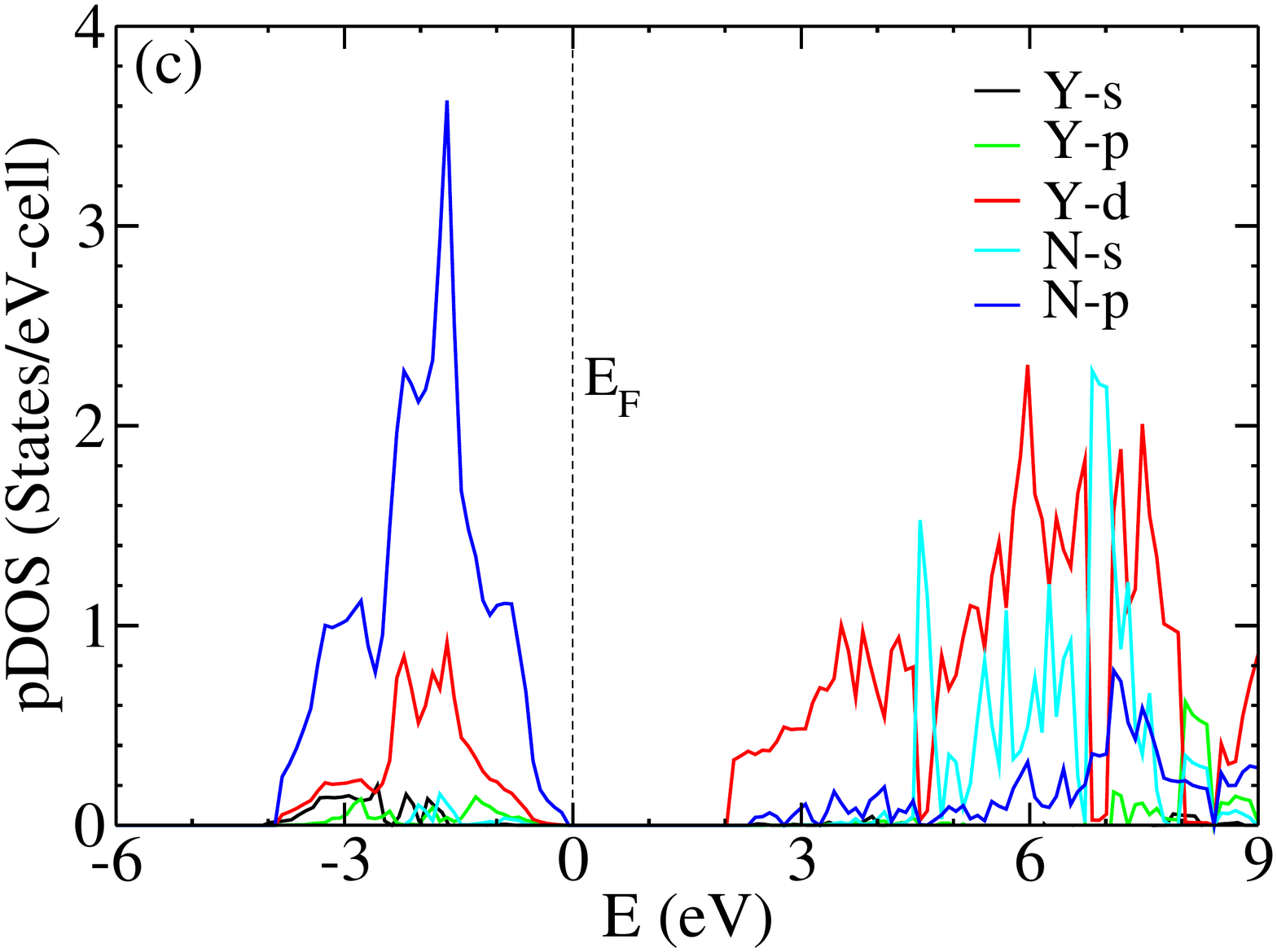}}}
    \end{minipage}
\end{center}
\caption{(Color online). (a) The calculated band structure of c-YN as obtained with the optimal basis set, at 
the equilibrium lattice constant of 4.827 \AA{}. 
The Fermi energy ($\mathrm{E_F}$) has been set equal to zero. 
(b) The calculated density of states (DOS) of c-YN as obtained from the bands shown in 
Fig.~\ref{fig:yn_elec}(a) The Fermi energy ($\mathrm{E_F}$) has been set equal to zero. 
(c) The calculated partial density of states (pDOS) of c-YN, 
as obtained from the bands (one with solid line) shown in Fig.~\ref{fig:yn_elec}(a). 
The Fermi energy ($\mathrm{E_F}$) has been set equal to zero.}
\label{fig:yn_elec}
\end{figure*}

Figures ~\ref{fig:scn_elec} to ~\ref{fig:yn_elec} exhibit the energy bands, and the related total 
(DOS) and partial (pDOS) densities of states of XN. 
All energies are referred relative to zero 
energy at the top of the valence band (VB). The electronic structures of the valence bands, the low energy conduction bands, 
and the band gap determine the most important properties of these materials in device applications.

Our ab-initio method shows that the fundamental gaps of both ScN and YN are indirect ones, with the maximum of 
the valence band ($VB_{max}$) occurring at $\Gamma$ and the minimum of the conduction band ($CB_{min}$) at the $X$ point 
(see Figs.~\ref{fig:scn_elec}(a) and ~\ref{fig:yn_elec}(a) for 
ScN and YN, respectively). Our calculated, indirect gaps for ScN, using the equilibrium lattice 
constant, is 0.79 eV and 0.88 eV for CA-VNW (LDA) and CA-PW (GGA) potentials, respectively. 
The difference in the band gaps and energy eigenvalues between 
CA-VWN and CA-PW may be attributed to the enhancement factor ($s$ $=$ $\frac{\lvert \triangledown n \rvert }{2 k_\mathrm{F}n}$) 
\cite{PhysRevB.54.16533,PhysRevB.64.180401} in the GGA functional. The band gap of 
ScN has been reported to suffer from the Burnstein-Moss shift  \cite{Burnstein1954,Moss1954} due to large background 
carrier concentration. As a consequence 
of this effect, several experimental indirect band gap values ranging from 0.90 $\pm$ 0.1 eV \cite{Brithen2004} 
to 1.32 $\pm$ 0.3 eV \cite{Gall2001} have been reported. 
Our computed band gap values are in good agreement with the lower, experimental values that 
result from relatively low carrier concentrations. For YN, our computations predict an 
indirect band gap value of 1.09 eV and 1.15 eV for LDA and GGA, respectively. To our knowledge, 
no measurement of the YN indirect gap has 
been reported. We expect YN to suffer from the same Burnstein-Moss effect as ScN, due to the 
close similarity of their structures.

Our calculated valence bands for ScN (see. Fig.~\ref{fig:scn_elec}(a)) and 
YN (see Fig.~\ref{fig:yn_elec}(a)) resemble those from other calculations available in literature. 
A major difference is the size of the band gap. For both ScN and YN, there is significant 
N$_{2p}$ -- X$_{id}$ (X = Sc, Y; and i = 3, 4 for Sc and Y, respectively) hybridization in the valence bands. 
The low energy conduction bands up to about 5.24 eV are mainly from X$_{id}$ 
(see Figs.~\ref{fig:scn_elec}(b) and ~\ref{fig:yn_elec}(b)  for ScN and YN, respectively). The following can 
further be confirmed from the partial 
density of states (see Figs.~\ref{fig:scn_elec}(c) and ~\ref{fig:yn_elec}(c) for ScN and YN, respectively). 
The upper valence bands originated 
from the bonding between N$_{2p}$ states and an admixture of 3$d$ states (for Sc) and 4$d$ states (for Y). 
The lowest conduction bands are mainly the antibonding t$_\mathrm{2g}$ states with 3d (for Sc) and 4$d$ (for Y) character, 
respectively. The two nonbonding $e_{g}$ bands are higher in energy. These bonding characters are in perfect agreement 
with the bonding analysis of \citet{Harrison1987}. The authors showed that rock-salt XN have 3 $p$-like bonding, 
3 $d$-like antibonding t$_\mathrm{2g}$, and two $d$-like nonbonding e$_\mathrm{g}$ 
bands formed by the hybridization of three valence $p$ 
states of N with the five $d$ states of X. The assignment of the bondings is consistent with the partial density 
of states analysis of the screened exchange LDA FLAPW calculations of Stampfl and co-workers 
\cite{Stampfl2001,PhysRevB.65.161204} and of the 
exact exchange based quasiparticle calculations of \citet{Qteish2006}, respectively.

The density of states (DOS) of ScN is shown in Fig. ~\ref{fig:scn_elec}(b).  In the valence bands, we found sharp peaks at 
--2.46 $\pm$ 0.1 eV and --2.96 $\pm$ 0.1 eV, respectively. A shallow minimum is observed at --3.05 $\pm$ 0.1 eV, 
followed by a shoulder at --3.2 $\pm$ 0.1 eV. A broad peak is located at --4.05 $\pm$ 0.2 eV. These peaks are 
formed by a strong hybridization between N $p$ states and Sc $d$ states with little contribution from 
Sc $p$ and $s$ states, respectively. In the conduction bands, the first observed, significant 
feature is a shoulder at about 3.80 $\pm$ 0.1 eV. It is due to the hybridization between N $p$ and 
Sc $d$ states. A sharp peak is observed at 4.7 $\pm$ 0.1 eV; it is formed from all the 
states of the atomic species except N $s$ states. The density of states of YN (see Fig.~\ref{fig:yn_elec}(b)) is similar 
to that of ScN, except that it has a pronounced shoulder at 0.92 eV below the Fermi energy. 
In the DOS of the valence bands, we predicted a sharp peak at --1.72 $\pm$ 0.1 eV, followed by two smaller ones at 
--2.1 $\pm$ 0.2 eV and --2.82 $\pm$ 0.1 eV, respectively. These peaks are mainly of N $p$ states and Y $d$ states. 
In the conduction bands, we observed a small peak at 3.6 $\pm$ 0.1 eV, which is 
mainly of N $p$ states and Y $d$ states. A sharp peak is observed at 5.94 $\pm$ 0.1 eV. This peak is 
of N $s$, N $p$, and Y $d$ states character.

To determine some specific properties of XN relevant to transport, we calculated the effective mass 
in different directions with respect to the $\Gamma$ point. The effective masses are 
calculated from curves that fit the bands in the immediate vicinity of minima (for electrons) or a 
maxima (for holes). The calculated electron effective masses (in units of $m_{o}$) at 
the bottom of the conduction band, along 
the $\Gamma$-$L$, $\Gamma$-$X$, and $\Gamma$-$K$ directions, respectively, are 0.91-1.66, 
0.78-0.99, and 0.87-1.07 for ScN, while those 
for YN are 0.89-1.02, 0.57-0.61, and 0.64-0.68 in the same direction as for ScN. 

\subsection{Chemical Bonding and Electron Charge Distribution}
\label{subsec:dos}
\begin{figure}
\centering
\includegraphics*[trim = 0mm 50mm 0mm 10mm,width=1\columnwidth,clip=true]{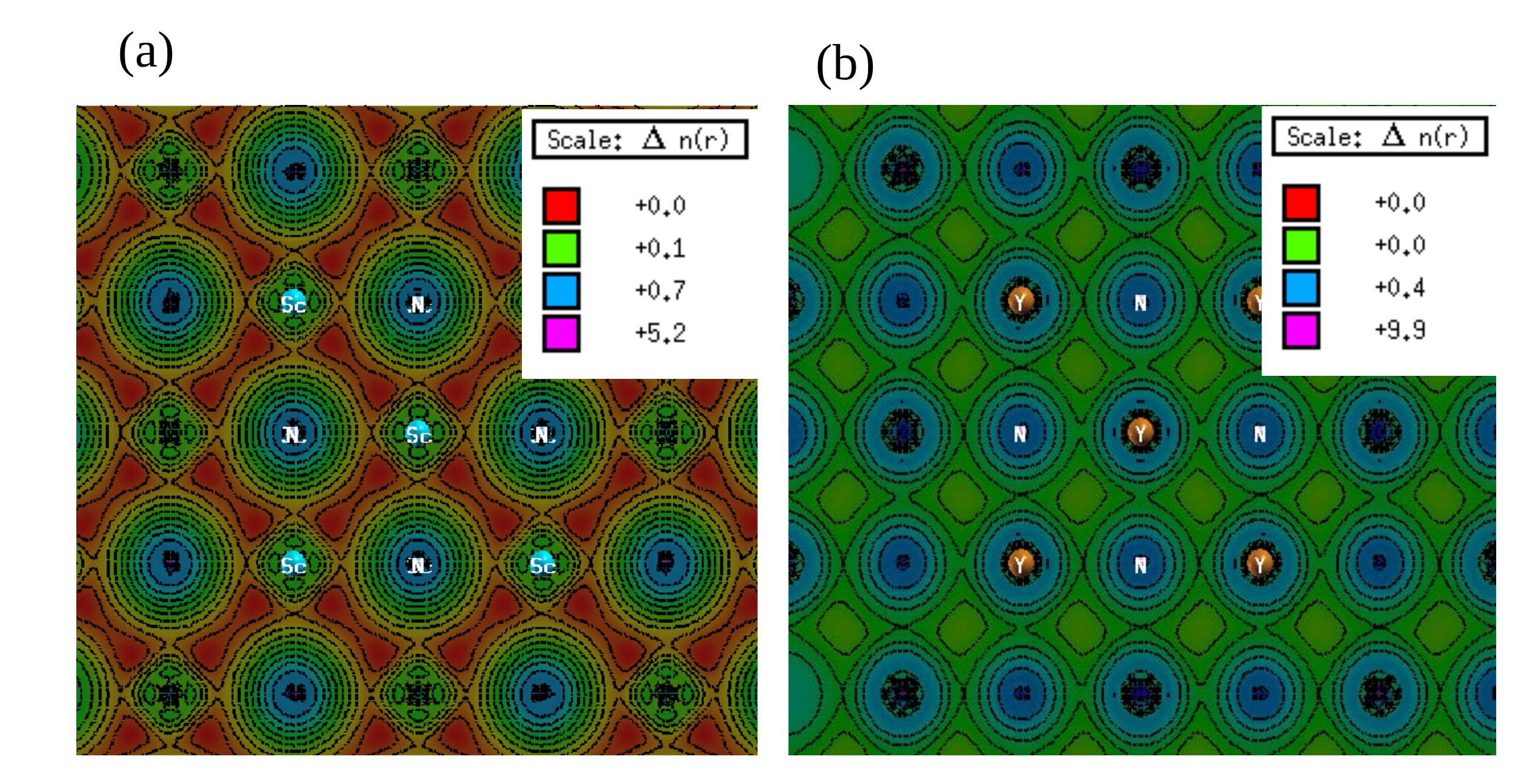}
\caption{(Color online) (a) The contour plot of the electron charge density of ScN as obtained with 
the optimal basis set. (b) The contour plot of the electron charge density of YN as obtained with 
the optimal basis set. $\Delta$ $n(r)$ is the variation of the 
electron charge density as a function of distance away from an atomic site. A logarithmic scale 
has been used.} 
\label{fig:xn_bond}
\end{figure}

Figure~\ref{fig:xn_bond} shows the contour plots of the 
distribution of the electron charge densities of XN along the (100) plane, cutting through the atoms. 
Away from the atomic centers (i.e., in the interstitial region) the electron charge density distributions are not 
spherically symmetric. The chemical bonding in 
ScN and YN appears to be of intermediate character between ionic bonding in the nonconducting 
calcium nitride or strontium nitride and that of the 
metallic bonding prevailing in the conducting group IV transition nitrides. As a consequence of this complex bonding, 
there seems to be a cooperative/competing bonding mechanisms in XN. As can be seen from 
Fig.~\ref{fig:xn_bond}, the bonding between X and N is covalent, with a significant ionic character. 
The bond length of Sc -- N is 2.25 \AA{} while that of Y -- N is 2.42 \AA{}. The experimental Sc -- N bond length is 
2.24 \AA{} \cite{gall:6034,ICSD2011} while the experimental bond length, Y -- N, is 
2.44 \AA{} \cite{Hayashi2003}. The covalent bonding in ScN 
can be seen to be stronger than that in YN. This difference is understandable in light of the larger Y -- N bond. 

\begin{table*}[htb] 
\centering
\caption{Calculated equilibrium lattice parameters a$_o$, stiffness, compliance and moduli constants, Young's modulus, 
Poisson ratio, shear modulus, and isotropic bulk modulus of ScN and YN.}
\begin{tabular}{cccccccccccccccccccccccccccc}
\cline{1-26}
\hline\hline
Nitride & a$_o$(\AA{}) &  & \multicolumn{3}{c}{C$_{ij}$ (GPa)} & \multicolumn{1}{c}{} & 
\multicolumn{4}{c}{$\nu$$_{\langle ijk \rangle}$} & \multicolumn{1}{c}{} & \multicolumn{3}{c}{C$_{ij}$ (10$^{-3}$ GPa$^{-1}$)} & 
\multicolumn{1}{c}{} & \multicolumn{4}{c}{E$_{\langle ijk \rangle}$ (GPa)} & \multicolumn{1}{c}{} & 
\multicolumn{3}{c}{$G$ (GPa)} & \multicolumn{1}{c}{} & $K$ &  &  \\ 
\cline{2-2}\cline{4-6}\cline{8-11}\cline{13-15}\cline{17-20}\cline{22-24}\cline{26-26}
Material &  &  & C$_{11}$ & C$_{12}$  & C$_{44}$  &  & $\nu$$_{\langle 100 \rangle}$ & 
$\nu$$_{\langle 110 \rangle}$ & $\nu$$_{\langle 111 \rangle}$ & $\bar{\nu}$ &  & S$_{11}$ & S$_{12}$ & S$_{44}$ &  
& E$_{\langle 100 \rangle}$ & E$_{\langle 110 \rangle}$ & E$_{\langle 111 \rangle}$ & $\bar{E}$ &  & $G_V$ & $G_R$ & $G$ &  &  &  &  \\ 
\cline{1-26}
ScN & 4.50 &  & 453 & 99 & 185 &  & 0.18 & 0.17 & 0.17 & 0.17 &  & 2.40 & -0.43 & 5.42  &  & 417 & 428 & 431 & 426 &  & 182 & 181 & 181 &  & 217 &  &  \\ 
YN & 4.83 &  &  405 & 93 & 145 &  & 0.19 & 0.20 & 0.20 & 0.20 &  & 2.70 & -0.50 & 6.88  &  & 371 & 355 & 350  & 358 & & 150 & 149 & 150 &  & 197 &  &  \\ 
\cline{1-26}
\hline\hline
\end{tabular}
\label{table:XN_elastic}
\end{table*}

\subsection{Structural and Elastic Properties}
\label{subsec:structural}

\begin{figure}
\centering
\includegraphics*[trim = 0mm 10mm 0mm 25mm, clip,totalheight=0.35\textheight, width=3.3in]{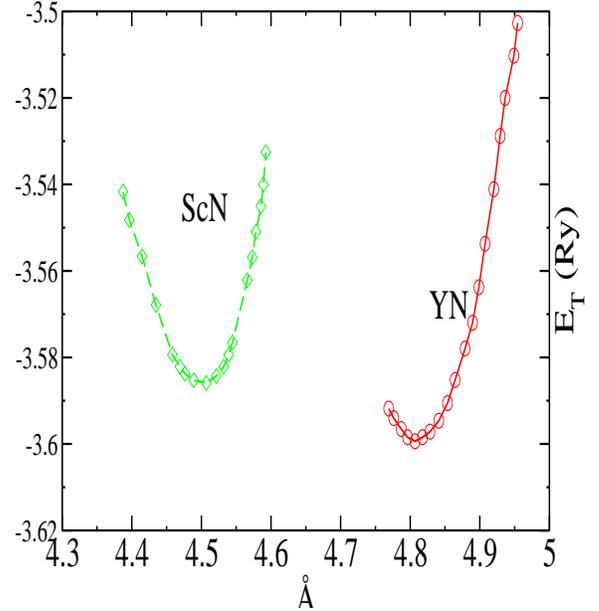}
\caption{(Color online) The total energy (E$_\mathrm{T}$) per unit cell as a function of lattice 
constants (\AA{}) for c-XN. Our calculated equilibrium lattice constant for ScN 
is 4.501 \AA{} (exactly the same as the experimental value). 
The calculated equilibrium lattice constant of YN is 4.827 \AA{}, 0.21 \% smaller than the experimental 
value of 4.837 \AA{}.} 
\label{fig:xn_eos}
\end{figure}

Total energy versus lattice constant data are as shown in Fig.~\ref{fig:xn_eos}  for ScN and YN. The computation was done 
with both the CA-VWN and CA-PW potentials. However, the reported data are for CA-PW, 
as we noted that the total energy difference between results from the two potentials 
were small ($\pm$ 0.12 eV). The data are fitted to 
the Murnaghan equation of state (EOS) for ScN and YN, respectively. 
The fit is very good, indicating that a different choice of EOS and/or lattice constant will have insignificant effect 
on the results.

For ScN, the calculated equilibrium lattice constant is 4.501 \AA{}, 
the bulk modulus ($B_{o}$) is 200.19 -- 216.73 GPa, and its pressure derivative ($B_{o}^\prime$) is 3.32 GPa. 
Experimental values are 4.501 \AA{} \cite{gall:6034,ICSD2011,gall:5524}, 182 $\pm$ 40 GPa \cite{gall:6034}, 
and 3.31 -- 3.89 GPa for 
the lattice constant, bulk modulus, and pressure derivative, respectively. For YN, our 
computations show that the equilibrium lattice constant is 4.827 \AA{}, the bulk modulus 
($B_{o}$) is 166.99 GPa. Our calculated equilibrium lattice constant of 4.827 \AA{} is practically the same as the 
experimental value of 4.837 \AA{}; it is just 0.21 \% smaller than the experimental value. 

Using the elastic constant relations as in Ref.~\cite{kim:1805}, 
we calculated the independent elastic 
constants C$_{11}$, C$_{12}$, and C$_{44}$. The calculated independent elastic constants is used to 
determine the compliance constants S$_{ij}$ (S$_{11}$, S$_{12}$, and S$_{44}$) \cite{kim:1805,Nye1957}. The Poisson ratios 
$\nu$ are measure of materials tendency to react to applied strain. For a cubic system (averaged over 
transverse directions in the [100], [110], and [111] directions), they are given by: \cite{Wojciechowski2004,PhysRevB.85.184106} 
$\nu_{\langle 100 \rangle}$ = -- $\frac{S_{12}}{S_{11}}$, $\nu_{\langle 110 \rangle}$ = 
-- $\frac{S_{11} + 2S_{12} - S_{44}/2)}{S_{11} + 2S_{12} + S_{44}}$, and $\nu_{\langle 111 \rangle}$ = 
-- $\frac{S_{11} + 3S_{12} - S_{44}/2}{2S_{11} + 2S_{12} + S_{44}}$. The isotropic shear modulus $G$ 
\cite{Schreiber1973} can be 
estimated from the average of the shear modulus in the Voigt approximation  ($G_V$ $=$ 1/5(C$_{11}$ -- C$_{12}$ + 3C$_{44}$)) 
\cite{Voigt1928} and Reuss approximation ($G_R$ $=$ $\frac{5 C_{44} (C_{11} - C_{12})}{4 C_{44} + 3 (C_{11} - C_{12})}$). 
\cite{Reuss1929} It is given as $G$ $=$ ($G_V$ + $G_R$)/2. Note that the Voigt and Reus approximations represent 
the upper and lower bounds of the isotropic shear modulus. In a typical cubic system, the isotropic 
bulk modulus $K$, can be calculated using the relation $K$ $=$ $\frac{1}{3} (C_{11} + 2 C_{12}$). 
Using these quantites, we obtain the isotropic Young's modulus, 
$\bar{E}$ = $\frac{9 K G}{3 K + G}$ and the isotropic Poisson ratio, $\bar{\nu}$ $=$ $\frac{3 K - 2 G}{2 (3 K + G)}$
$\approx$ $\frac{1}{3}$($\nu_{\langle 100 \rangle}$ + $\nu_{\langle 110 \rangle}$ +$\nu_{\langle 111 \rangle}$). 
The equilibrium elastic constants, compliance constants 
(S$_{11}$, S$_{12}$, and S$_{44}$), the elastic moduli corresponding to the determined
stiffness constants and other elastic parameters are shown in Table~\ref{table:XN_elastic}. 

The calculated,  
independent elastic constants for ScN are C$_\mathrm{11}$ $=$ 452.55 GPa, 
C$_\mathrm{12}$ $=$ 98.82 GPa, and C$_\mathrm{44}$ $=$ 184.60 GPa. 
Using the Zener's anistropy index (ratio) \cite{Zener1948,PhysRevLett.101.055504}, $\eta$ $=$ 2C$_{44}$/(C$_{11}$ - C$_{12}$), 
we quantify the anistropy as 1.04. This quantity contains the same information as the ratio of the directional Young's 
modulus (E$_\xi$, $\xi$ is direction) E$_{\langle 111 \rangle}$/E$_{\langle 100 \rangle}$, suggesting that ScN is stiffer 
in the ${\langle 111 \rangle}$ than ${\langle 100 \rangle}$ direction. The elastic moduli ratio 
E$_{\langle 100 \rangle}$:E$_{\langle 110 \rangle}$:E$_{\langle 111 \rangle}$ is 1.00:1.03:1.03.
For YN, the calculated, independent elastic constants are: 
C$_\mathrm{11}$ $=$ 405 GPa, C$_\mathrm{12}$ $=$ 92.70 GPa, and C$_\mathrm{44}$ $=$ 145.28 GPa. The 
anisotropy ratio $\eta$ is 0.93. The elastic moduli ratio is 
E$_{\langle 100 \rangle}$:E$_{\langle 110 \rangle}$:E$_{\langle 111 \rangle}$ = 1.0:0.96:0.94. 

For ScN, $\bar{E}$ $=$ 425.68 GPa, $G$ $=$ 181.47 GPa, 
and $\bar{\nu}$ = 0.17. Our computed results are comparable to experimental results of \citet{gall:6034}, 
\citet{moram:023514}, and \citet{gall:5524} on 
epitaxial films of ScN. These authors reported values of 
$E$ in the range 270 $\pm$ 25 to 388 $\pm$ 20 GPa and $v$ value of 0.15, 0.188 $\pm$ 0.002 and 0.20 $\pm$ 0.04. Our calculated and 
measured $\bar{\nu}$ values of ScN are in the same range. Our calculated value of $E$ is close to the upper bound, given above, 
of the measured values for the films. The slight 
differences between some of our computed elastic constants and experimental ones may be due to the fact that these 
experiments were done on films of ScN. We should note that growth conditions and film thickness \cite{gall:6034} are known to 
affect significantly the elastic 
constants, as is evident in the works of \citet{gall:6034}, \citet{gall:5524}, and \citet{moram:023514} which showed 
a wide range of values.   
For YN, $\bar{E}$ $=$ 357.94 GPa, $G$ $=$ 149.53 GPa, 
and $\bar{\nu}$ = 0.20. We note that our computed Poisson ratios, $\bar{\nu}$ are comparable to known Poisson 
ratios for other transition-metal nitrides that are in the range 0.19 -- 0.22 \cite{Sue1992,T1987,kim:1805}. 
We are not aware of any experimental measurements of $E$, $G$, and $v$ for YN; hence, our calculated results are predictions. 
In both ScN and YN, our computed 
isotropic bulk modulus, $K$ is in basic agreement with the bulk modulus we obtained for the equilibrium structural optimization. 

Our calculated elastic stiffness constants, obey the 
stability conditions for a cubic crystal system, i.e., 
B$_o$ $=$ (C$_{11}$ + 2C$_{12}$)/3 $>$ 0, $G'$ $=$ (C$_{11}$ - C$_{12}$)/2 $>$ 0, and 
C$_{44}$  $>$ 0 \cite{PhysRevB.60.R8449}. The calculated $\eta$ values are 
both close to 1, implying that the elastic behavior is almost 
isotropic. The Cauchy pressure (C$_{12}$ - C$_{44}$), in both 
ScN and YN, is significantly less than 0. It corresponds to a directional bonding 
\cite{chen:2414,Pettifor1992,PhysRevB.85.064101}, leading 
to large charge transfers from cations to anions as has been observed experimentally in TiN \cite{PhysRevB.83.165122}.
  
\section{Summary and Conclusion}
\label{sec:summary}
We have performed first principle calculations of electronic and related properties of ScN 
and YN, based on basis set optimization with the 
BZW-EF method. Accurate, band gaps, structural parameters, lattice constants, bulk moduli, and elastic 
constants were calculated. We have provided detailed analyses of the electronic and related properties of both ScN and YN. 
Our computed properties 
are in good agreement with experiment. In particular, we predict various electronic (indirect band gap value), 
effective masses, elastic and structural properties of YN. It is our hope that the present results will motivate further experimental 
and theoretical studies of the properties of these materials, with emphasis on measurements of electronic and elastic 
properties of YN.  

\begin{acknowledgments}
High performance computational 
resources is provided by Louisiana Optical Network Initiative (LONI). 
This work was funded in part by the the National Science Foundation 
(Award Nos.: EPS-1003897 and NSF (2010-15)-RII-SUBR, and HRD 100254). 
The Louisiana Space Consortium (LaSPACE, Sub-award No. 56726), and LONI-SUBR. 
CEE wishes to thank Govt. of Ebonyi State, Nigeria.
\end{acknowledgments}


\end{document}